\documentclass[conference]{IEEEtran}
\IEEEoverridecommandlockouts
\usepackage{cite}
\usepackage{amsmath,amssymb,amsfonts}
\usepackage{algorithm}
\usepackage{algpseudocode}
\usepackage{graphicx}
\usepackage{textcomp}
\usepackage{xcolor}
\usepackage{booktabs}
\usepackage{url}
\usepackage{multirow}
\usepackage{tikz}
\usepackage{makecell}
\usepackage{placeins}
\usepackage{setspace}
\usepackage[hidelinks]{hyperref}
\usepackage{array}
\usetikzlibrary{arrows.meta, positioning, calc, backgrounds}

\def\BibTeX{{\rm B\kern-.05em{\sc i\kern-.025em b}\kern-.08em
    T\kern-.1667em\lower.7ex\hbox{E}\kern-.125emX}}

\algnewcommand\algorithmicinput{\textbf{Input:}}
\algnewcommand\algorithmicoutput{\textbf{Output:}}
\algnewcommand\Input{\item[\algorithmicinput]}
\algnewcommand\Output{\item[\algorithmicoutput]}

\begin{document}
\begin{titlepage}
\vspace*{\fill}
\begin{center}
\textbf{Notice}\\[1em]
\small
© 2026 IEEE. Personal use of this material is permitted.
Permission from IEEE must be obtained for all other uses,
in any current or future media, including
reprinting/republishing this material for advertising or
promotional purposes, creating new collective works, for
resale or redistribution to servers or lists, or reuse of
any copyrighted component of this work in other works.\\[2em]
This is the author's accepted manuscript version of the
following paper:\\[1em]
\textbf{AutoTrans: AI-Assisted Automatic Translation of
Security Assertions for RISC-V Processors}\\[0.5em]
Sharjeel Imtiaz, Uljana Reinsalu, Tara Ghasempouri\\[0.5em]
Accepted at IEEE Baltic Electronics Conference (BEC) 2026.\\[0.5em]
The final published version will be available on IEEE Xplore.
\end{center}
\vspace*{\fill}
\end{titlepage}

\setstretch{0.93}
\title{AutoTrans: AI-Assisted Automatic Translation of Security
Assertions for RISC-V Processors%
\thanks{\scriptsize This work is supported by the Estonian
Research Council grant PSG837, the Estonian--French
PARROT program, and the EU Horizon Europe Grant Agreement
No.\,101160182 TAICHIP.}}

\author{
\IEEEauthorblockN{Sharjeel Imtiaz, Uljana Reinsalu, Tara Ghasempouri}
\IEEEauthorblockA{
  \textit{Department of Computer Systems,
  Tallinn University of Technology,
  Tallinn, Estonia} \\
  \{sharjeel.imtiaz, uljana.reinsalu, tara.ghasempouri\}@taltech.ee}
}

\maketitle

\begin{abstract}
Reusing a set of verified security assertions across RISC-V processor targets remains one of the most expensive bottlenecks in hardware security verification. Manual translation takes hours per assertion. Raw LLM translation is fast but unreliable, introducing signal hallucination, where the model invents port names absent from the target RTL and produces outputs that may vary across model updates or even within the same model version. This paper presents AutoTrans, an automated framework that addresses the above shortcomings. First, a new lightweight Regular Expression-based System Verilog signal extractor is proposed to identify the signals for generating security assertions. This step is necessary to prevent signal hallucination. Second, a template is introduced to create prompt and pinned inference parameters that guarantee a byte-identical prompt assembly on every run, making the pipeline output resilient to model updates. Moreover, the introduced template for LLM prompting is capable of generating security assertions from English-only security descriptions of RISC-V processors, with no manual authoring. Third, a formal verification process (JasperGold FPV) is integrated, which guarantees that the generated security assertions verify the security of the RISC-V processor rather than silently entering the result set. The workflow is applied on Deepseek V4 to translate security assertions from one RISC-V to another (e.g., for IBEX from NS31A RISC-V). The experiment shows that AutoTrans achieves 78\% Auto Translation Acceptance Rate (TAR) automatically and without human intervention and 100\% Final TAR after refinement by humans.
\end{abstract}


\section{Introduction}
\label{sec:intro}

Security assertion development is one of the most
time-intensive and expertise-demanding tasks in hardware
security verification; encoding a single
privilege violation, invariant, or memory protection
rule~\cite{RISC-V-ISA-p,RISC-V-ISA-up} requires hours of
RTL analysis per assertion, and a complete processor corpus
runs to weeks of effort. Yet when a new processor target is adopted, this investment is discarded. RISC-V has accelerated this problem; the ISA has produced dozens of implementations
sharing the same instruction set but diverging significantly
in microarchitecture~\cite{riscv_security_survey,
riscv_security_survey2}, and security knowledge should travel with the ISA, but currently does not. For instance, the NS31A~\cite{ns31a} corpus of
1146 security property instances, grouped into 68 distinct
assertion groups across nine categories, was manually authored
for a reference RISC-V processor with significant RTL expertise
per assertion, and must be entirely reworked to port to a
structurally different target such as lowRISC
Ibex~\cite{ibex}: every signal name, bit width, addressing mode,
and clocking convention must be matched to the new RTL, a process
measured in hours per assertion and weeks per corpus.

Three classes of prior work address parts of this problem, but none solve it. All three share the same fundamental limitation: they cannot bridge semantic gaps between
architecturally heterogeneous designs while guaranteeing formal correctness.
The first class, NLP-based approaches~\cite{soeken2014},
use dependency parsing and handcrafted templates to
translate natural-language specifications into formal
properties. They perform well within a single design
family, but break down when signal semantics differ
fundamentally across architectures.
The second class, structural translation tools, is
represented by Transys~\cite{transys}, which automates
security assertion translation using Program Dependence
Graphs and achieves strong results on syntactically
similar designs. However, when no structural counterpart
exists in the target RTL, the dependency graph has
nothing to match. For example, the source processor
corpus assumes unconditional debug-mode bypass while
Ibex restricts it to a specific Debug Module address
range  a semantic gap that structural graph matching
cannot resolve.
The third class, LLM-based generation
approaches~\cite{autoassert,assertllm,lasa,divas,chauhan},
produce new assertions from RTL or natural language.
They target a fundamentally different problem: creating
new claims rather than porting verified ones. Their
outputs are evaluated with BLEU, a text similarity
score, or coverage metrics, neither of which confirms
formal correctness. These approaches additionally require
GPU clusters for fine-tuning or inference, making them
inaccessible for standard verification workflows.

Raw LLM translation introduces two failure modes that prior work does not address. Without knowledge of the target RTL interface, the LLM hallucinates signal names absent from the RTL and wraps output in Markdown fences, both causing immediate QuestaSim compilation failure
(Section~\ref{sec:ablation}). And even with correct output, cloud providers may silently update model weights, producing structurally different bind files from identical
inputs with no warning. Solving both requires three things simultaneously: deterministic signal grounding to
prevent hallucination, a fixed prompt structure for reproducible assembly, and a formal FPV gate to catch any remaining variation before it enters the result set.

AutoTrans addresses this gap directly. Prior
work~\cite{iscas} demonstrated semi-automated SVA-to-SVA
translation on five modules; this paper delivers a fully
automated pipeline across all nine Ibex security modules
with a formal verification gate at every step. The contributions of this work are:

\begin{itemize}
    
      \item Formally verified security assertions authored
          for one RISC-V processor are automatically
          translated to a structurally different target
          through the AutoTrans pipeline, in which every
          assertion must pass the QuestaSim compilation and
          JasperGold FPV, preserving security intent
          without manual re-authoring, GPU, or fine-tuning.

    \item A lightweight Regular Expression-based signal
        extractor is proposed that parses the target RTL
        interface and grounds every LLM prompt in the
        exact available signals, addressing hallucination
        without requiring EDA infrastructure or licenses.

   \item A fixed prompt template combined with pinned
        inference parameters enforce byte-identical
        prompt assembly on every run, enabling fully
        automated and reproducible translation with
        no GPU and no fine-tuning required.
        
    \item The Translation Acceptance Rate (TAR) is
        introduced as a per-module metric that counts
        only assertions formally proven correct and
        non-vacuous on the target RTL by industry-grade
        verification tools.

\end{itemize}

Applied to 68 security property groups from a reference
RISC-V processor across nine Ibex modules, AutoTrans
achieves \textbf{78\% Auto TAR} and \textbf{100\% Final TAR}.

The remainder of this paper is organized as follows.
Section~\ref{sec:related} surveys related work.
Section~\ref{sec:framework} describes the AutoTrans
framework. Section~\ref{sec:experiments} presents
experimental results. Section~\ref{sec:conclusion}
concludes.

\section{Related Work}
\label{sec:related}

\subsection{Assertion-Based Security Verification}

A corpus of 1146 security properties across nine categories
was manually authored for a reference RISC-V processor,
requiring significant RTL expertise per
assertion~\cite{ns31a}. Hardware Trojans represent a
critical threat class in a modern processor
designs~\cite{bhunia2014}; security properties for
open-source RISC-V designs have been shown to detect them
alongside privilege violations and illegal memory
accesses~\cite{heidari2025,norcasreza,iscas}. All of these
works depend on correct, target-specific assertions being
available. AutoTrans removes this bottleneck by delivering
formally verified, automatically translated assertions
without manual porting or ground-up generation.

\subsection{Non-LLM Assertion Translation}

Early work applied NLP dependency parsing and handcrafted
templates to convert natural-language specifications into
formal properties~\cite{soeken2014}. This approach works
within a single design family where assertion patterns are
uniform but does not scale to architecturally heterogeneous
targets where signal semantics differ fundamentally between
source and destination.

Transys~\cite{transys} is the only prior system that
automates security assertion translation across hardware
designs without an LLM. It extracts Program Dependence
Graphs (PDG) from the source and target RTL and applies
statistical feature matching and structural transformations
to produce translated properties. Transys achieves strong
results on syntactically similar designs, but requires EDA
tools to build the PDG and struggles when the source
assertion encodes a security model with no structural
counterpart in the target RTL. AutoTrans addresses precisely this class of semantic gap by using LLM reasoning grounded in the exact target
RTL interface.

\subsection{LLM-Based Assertion Generation}

Recent LLM work targets assertion \emph{generation} from
RTL or natural language rather than the translation of a
verified corpus. Fine-tuned approaches~\cite{autoassert,
assertllm} train on RTL corpora and evaluate output with BLEU/ROUGE requires GPU clusters for training.
Prompt-based approaches~\cite{lasa, divas,chauhan} use
general-purpose LLMs with iterative feedback or structured
templates, reporting accuracy on internal tests without a
formal verification gate. Template-mask
instantiation~\cite{sabry2026} confirms that fixed
structural templates are central to syntactically valid
assertion synthesis.

None of these works port an existing verified corpus to
a new target. BLEU and coverage metrics measure syntactic
proximity, not formal correctness on the target RTL.
AutoTrans requires every assertion to be formally proven
and non-vacuous on clean RTL as a hard acceptance gate; model drift can only cause FPV failure and retry, never a wrong assertion entering the result.

\begin{figure}[t]
\centering
\includegraphics[width=\columnwidth]{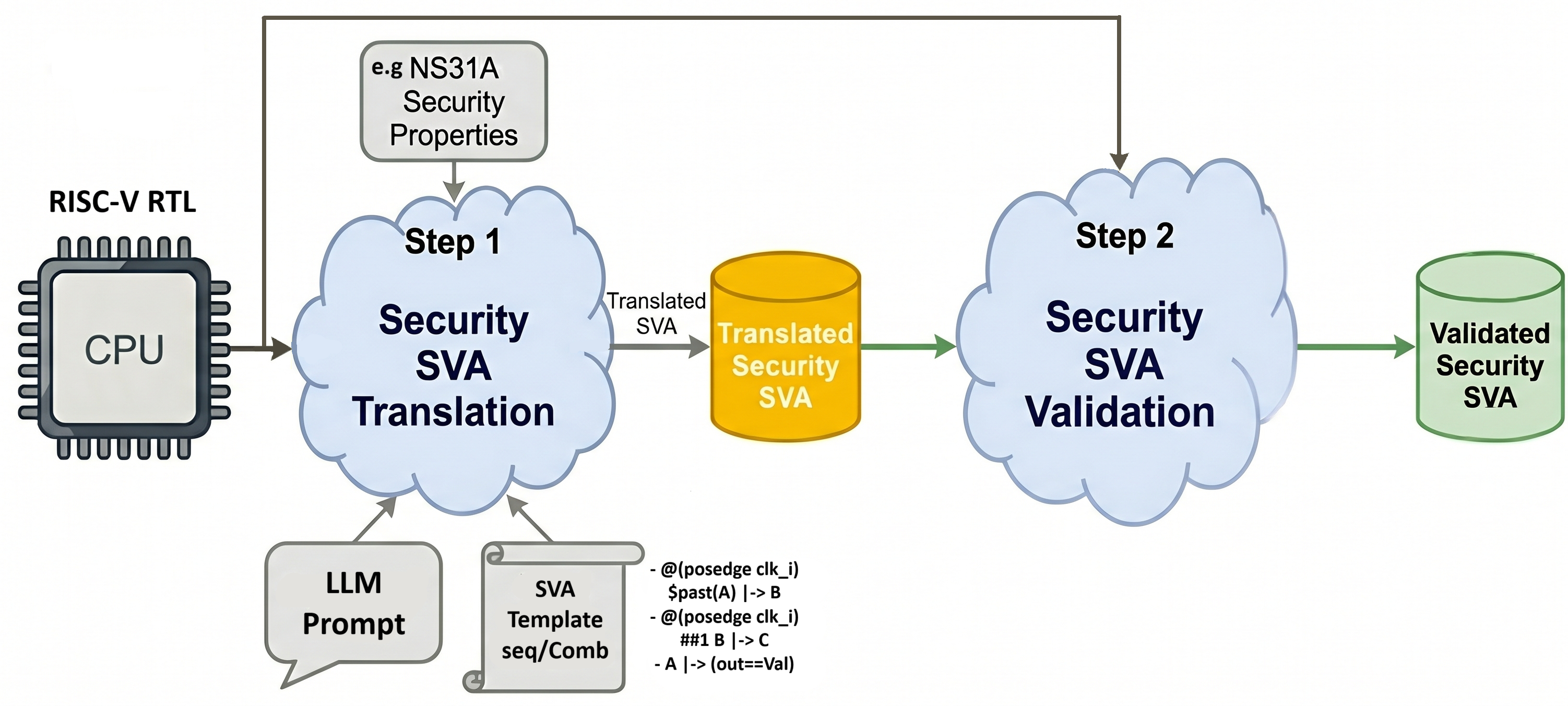}
\caption{AutoTrans two-stage framework overview.}
\label{fig:overview}
\end{figure}
 
\section{AutoTrans Framework}
\label{sec:framework}

\subsection{Overview}

The AutoTrans framework operates in two stages as shown in
Fig.~\ref{fig:overview}. Stage~1 (Security SVA Translation)
takes the target RISC-V RTL and security properties
from the source processor~\cite{ns31a} as input and produces translated SVA bind files.
Stage~2 (Security SVA Validation) takes those bind files
and produces formally validated assertions alongside a TAR
report (defined in Section~\ref{sec:tar}).
Algorithms~\ref{alg:extract} and~\ref{alg:prompt} formalize
Step~1A and Step~1B; Section~\ref{sec:stage2} covers validation.

\subsection{Stage~1: Security SVA Translation}
\label{sec:stage1}

\begin{figure}[H]
\centering
\includegraphics[width=\columnwidth]{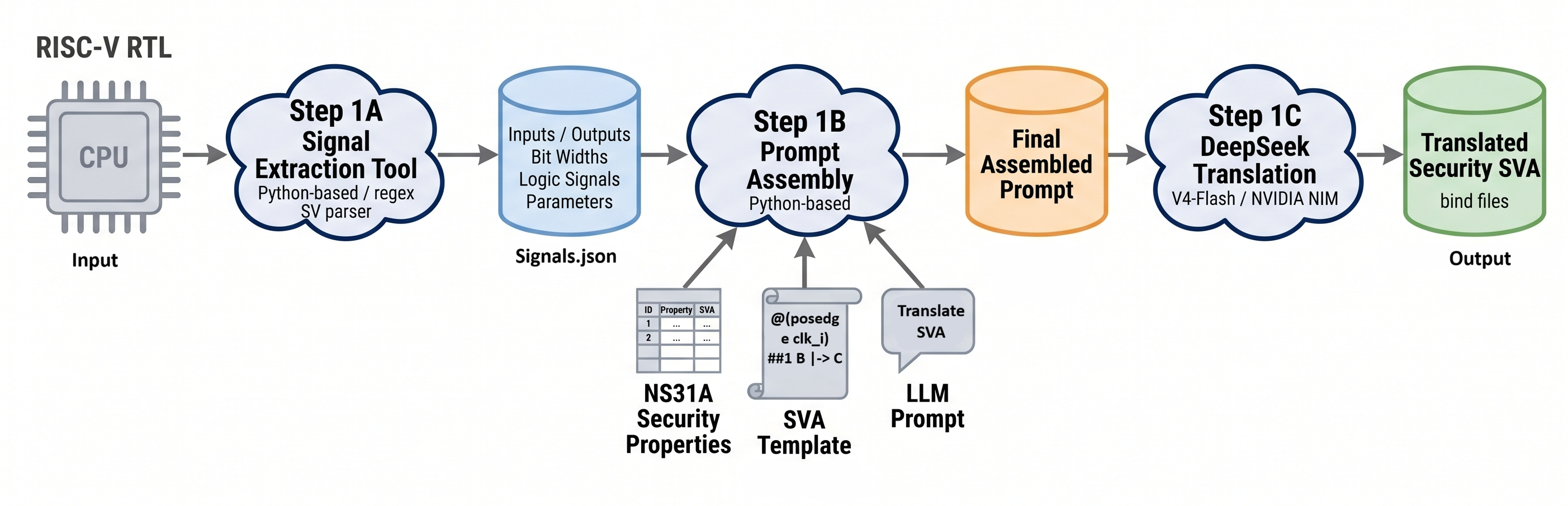}
\caption{Stage~1: signal extraction to \texttt{json file}
(Step~1A), grounded prompt assembly (Step~1B), and DeepSeek
V4-Flash SVA translation (Step~1C).}
\label{fig:step1}
\end{figure}

Stage~1 comprises three steps shown in Fig.~\ref{fig:step1}. Step~1A extracts signals from the target RTL as formalized in Algorithm~\ref{alg:extract}, Step~1B assembles the final prompt from those signals and the source processor security properties as formalized in Algorithm~\ref{alg:prompt}, and Step~1C calls DeepSeek V4-Flash to produce a bind file, a separate SystemVerilog module containing the translated assertions, attached to the target RTL at elaboration time without modifying it.

 \vspace{-12pt}
 \vspace{-8pt}

\subsubsection{Step~1A: RTL Signal Extraction}
\label{sec:extraction}

The signal extractor, formalized in Algorithm~\ref{alg:extract}
and illustrated in Fig.~\ref{fig:step1}, is a custom Regular Expression-based SystemVerilog parser implemented in Python. Existing tools were not designed
for this task, PyVerilog~\cite{pyverilog} targets Verilog-era
syntax, Slang~\cite{slang} and Verilator~\cite{verilator} are
C++ tools with no Python signal extraction interface, and
hdlparse~\cite{hdlparse} does not resolve package-qualified
types. None produces structured JSON for a direct LLM prompt
injection, which this pipeline requires.

Algorithm~\ref{alg:extract} proceeds in four stages.
If a package file is provided, it is parsed first (lines~1--8):
typedef enums yield symbolic constants and values (lines~2--4);
typedef structs yield field names and types (lines~5--7).
For each SV file (lines~9--23), the source is read and stripped
of comments and compiler directives (line~10); module type is
determined immediately from \texttt{always\_ff} presence
(lines~11--15); parameters are extracted (line~16); ANSI port
declarations are parsed with width resolved from explicit bounds,
package-qualified types, or scalar default (lines~17--19); and
internal signal declarations are extracted, excluding port names
and SV keywords (lines~20--22).
If multiple files are provided, ports, internals, and parameters
are merged with deduplication (lines~24--26).
Clock (\texttt{clk\_i}) and reset (\texttt{rst\_ni}) are detected
from port name conventions for sequential modules (lines~27--29).
Package types are filtered to those referenced in port and
internal widths (line~30), and all data is serialized to
\texttt{signals.json} (line~31), the sole input to downstream
prompt assembly.

 \vspace{-6pt}
\begin{algorithm}[H]
\scriptsize
\caption{RTL Signal Extraction (Step~1A)}
\label{alg:extract}
\begin{algorithmic}[1]
\Input{SV file(s), optional \texttt{ibex\_pkg.sv}}
\Output{\texttt{signals.json}}
\If{pkg file provided}
  \ForAll{typedef enum in pkg}
    \State Extract symbolic constants and values
  \EndFor
  \ForAll{typedef struct packed in pkg}
    \State Extract field names and types
  \EndFor
\EndIf
\ForAll{SV file(s)}
  \State Read file; strip comments and compiler directives (\texttt{`ifdef}, \texttt{`include})
  \If{\texttt{always\_ff} found} \State type $\leftarrow$ sequential
  \Else\ \State type $\leftarrow$ combinational \EndIf
  \State Extract parameter names (\texttt{parameter} / \texttt{localparam})
  \ForAll{port declarations in ANSI port list}
    \State Extract direction, name; resolve width from \texttt{[N-1:0]},
    \Statex \hspace{3.5em} package-qualified type, or scalar default (1~bit)
  \EndFor
  \ForAll{internal declarations (\texttt{logic}/\texttt{wire}/\texttt{reg}/typedef)}
    \State Extract name and width; skip port names and SV keywords
  \EndFor
\EndFor
\If{multiple files} \State Merge ports, internals, parameters with deduplication \EndIf
\If{sequential} \State Detect clock (\texttt{clk\_i}) and reset (\texttt{rst\_ni}) from port names \EndIf
\State Filter pkg types to those referenced in port/internal widths
\State Serialize all to \texttt{signals.json}
\end{algorithmic}
\end{algorithm}
 \vspace{-6pt}

\subsubsection{Step~1B: Prompt Assembly}
\label{sec:promptassembly}

Formalized in Algorithm~\ref{alg:prompt} and shown in
Fig.~\ref{fig:step1}, this step assembles the final prompt
from four components: the \texttt{signals.json} output from
Step~1A, the NS31A security properties from the CSV file,
the SVA template selected by module type (sequential or
combinational), and the LLM prompt template, which together
produce the final assembled prompt passed to Step~1C.

Algorithm~\ref{alg:prompt} proceeds in three stages.
First, inputs are loaded (lines~1--2): \texttt{signals.json}
provides module type, ports, internals, and package types; the source processor CSV provides the source assertion groups; the SVA
template and the LLM prompt template are selected in the following stage.
The template is then selected by module type (lines~3--7):
sequential modules load the clocked template, mandating
\texttt{@(posedge clk\_i) disable iff (!rst\_ni)},
\texttt{\#\#N}, and \texttt{\$past()}; the PMP module loads
the combinational template which forbids all clocking
constructs.
Finally, all placeholders are filled in a single pass
(lines~8--10): module name and shared-RTL logical suffix
for DO/ETI/CF/MT, clock and reset signals, port and
internal signal lists, package types formatted as SV
\texttt{typedef} text, port declarations with all DUT
ports declared as \texttt{input}, and targeted RISC-V assertion
groups formatted from CSV rows. The filled prompt is
saved to \texttt{prompts/final/} (line~10) for reproducibility auditing.

 \vspace{-6pt}
\begin{algorithm}[H]
\scriptsize
\caption{Prompt Assembly (Step~1B)}
\label{alg:prompt}
\begin{algorithmic}[1]
\Input{\texttt{signals.json}, NS31A CSV, SVA template, LLM prompt template}
\Output{\texttt{prompts/final/\textit{MODULE}\_final\_prompt.txt}}
\State Load \texttt{signals.json}; read module type, ports, internals, pkg\_types
\State Load NS31A CSV rows
\If{module type = sequential}
  \State Load sequential template (clocked, \texttt{disable iff})
\Else
  \State Load combinational template (clock-free)
\EndIf
\State Derive assertion module name; for shared-RTL modules (DO/ETI/CF/MT)
\Statex \hspace{1.5em} append logical suffix (e.g.\ \texttt{ibex\_controller\_do})
\State Fill all \texttt{\{\{PLACEHOLDERS\}\}} in one pass:
\Statex \hspace{1.5em} \texttt{\{\{MODULE\_NAME\}\}}, \texttt{\{\{MODULE\_SHORT\}\}},
                       \texttt{\{\{CLOCK\}\}}, \texttt{\{\{RESET\}\}}, \texttt{\{\{PARAMETERS\}\}}
\Statex \hspace{1.5em} \texttt{\{\{INPUT\_PORTS\}\}}, \texttt{\{\{OUTPUT\_PORTS\}\}},
                       \texttt{\{\{INTERNAL\_SIGNALS\}\}}
\Statex \hspace{1.5em} \texttt{\{\{PKG\_TYPES\}\}}: pkg\_types formatted as SV
                       \texttt{typedef} text (already resolved by Step~1A)
\Statex \hspace{1.5em} \texttt{\{\{PORT\_DECLARATIONS\}\}}: all DUT ports as
                       \texttt{input}; clock/reset excluded for sequential modules
\Statex \hspace{1.5em} \texttt{\{\{NS31A\_ASSERTIONS\}\}} formatted from CSV rows;
                       \texttt{\{\{NS31A\_TOTAL\_GROUPS\}\}} = CSV row count
\State Save filled prompt to
       \texttt{prompts/final/\textit{MODULE}\_final\_prompt.txt}
\end{algorithmic}
\end{algorithm}
 \vspace{-6pt}

\subsubsection{Step~1C: DeepSeek Translation}
\label{sec:translate}

As shown in Fig.~\ref{fig:step1}, the translation step is implemented in Python. The filled prompt is submitted to the NVIDIA NIM API~\cite{nvidiaNIM}, calling
DeepSeek V4-Flash (\texttt{DeepSeek-ai/DeepSeek-v4-flash})
with \texttt{temperature=0.0} and \texttt{seed=42}, which
minimize sampling variance through greedy decoding. The
two-tier strategy is a deliberate cost-efficiency decision:
V4-Flash handles initial translation at lower inference
cost, while V4-Pro is reserved for retry calls where
QuestaSim or JasperGold has identified a specific failure,
concentrating the higher-cost model only where deeper RTL
reasoning is needed. No GPU is required; NVIDIA NIM
provides 1,000 free developer credits, sufficient to cover
all nine modules. No fine-tuning is performed.

The \texttt{<think>} reasoning trace is stripped before
parsing; the remaining response yields a JSON mapping log
and a complete SVA bind file. The four modules sharing
\texttt{ibex\_controller.sv} (DO, ETI, CF, MT) have their
\texttt{bind} target rewritten to the correct RTL module
name. Every assertion group is translated without
exception; when an source RISC-V processor signal has no direct targeted processor
equivalent, the template instructs the model to preserve
security intent using the nearest available Ibex signals.
Translation coverage is always 100\% at Step~1C; TAR is
computed only after FPV acceptance at Step~2B.

\subsection{Stage~2: Security SVA Validation}
\label{sec:stage2}

\begin{figure}[H]
\centering
\includegraphics[width=\columnwidth]{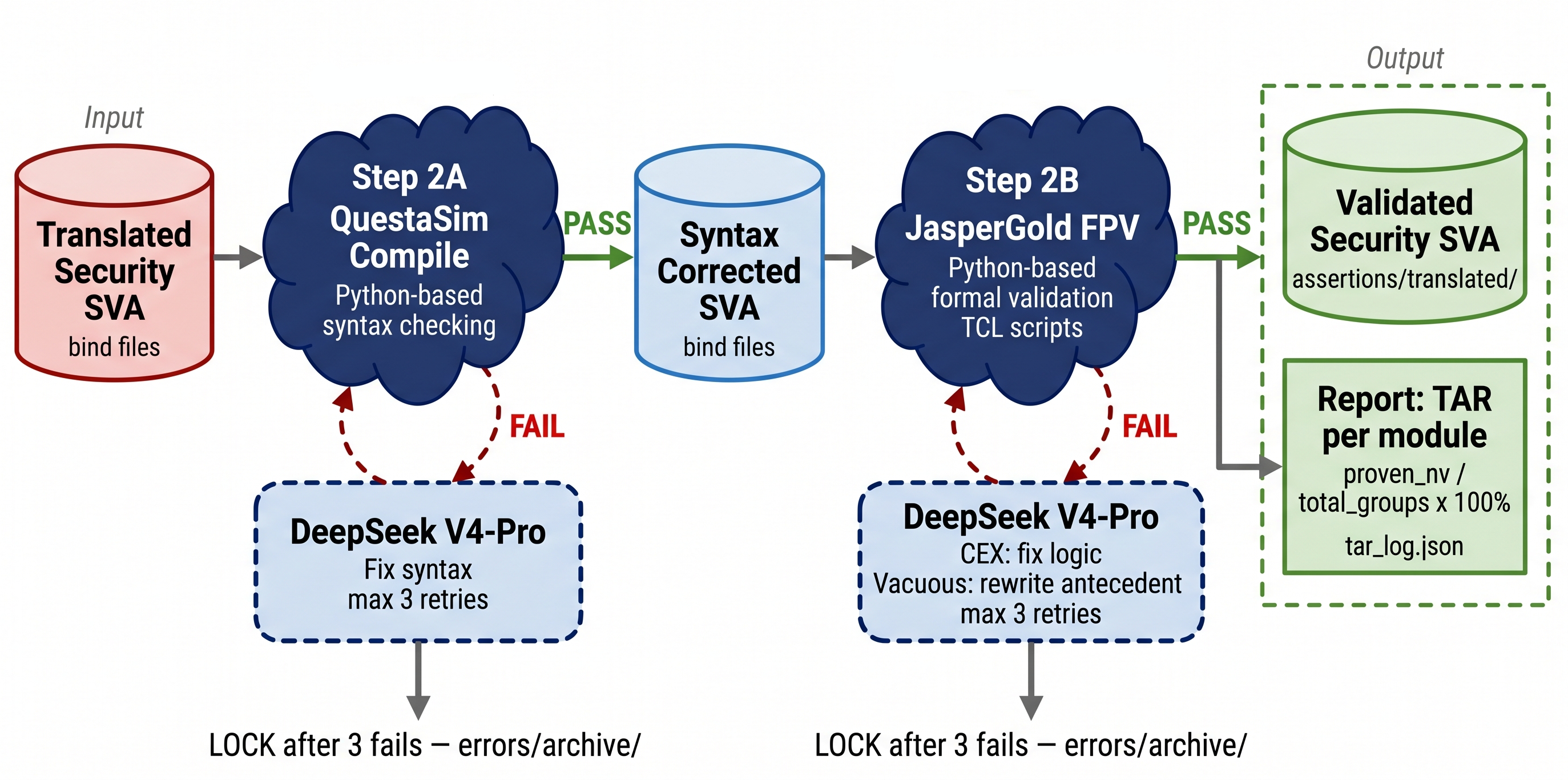}
\caption{Stage~2: QuestaSim syntax validation (Step~2A)
and JasperGold FPV verification (Step~2B) with V4-Pro
retries.}
\label{fig:step2}
\end{figure}

Stage~2 comprises two steps shown in Fig.~\ref{fig:step2}.
Step~2A compiles the translated bind files through
QuestaSim to catch syntax errors; Step~2B runs JasperGold
FPV to verify that every assertion is formally proven and
non-vacuous on clean RTL. This two-gate design is
deliberate: syntax errors require deep RTL reasoning to
fix, while FPV failures require understanding assertion
intent at the semantic level. V4-Pro handles all retries
at both steps with a maximum of three before the module
is locked.

 \vspace{-10pt}
 \vspace{-8pt}

\subsubsection{Step~2A: Syntax Validation}
The translated bind file is compiled together with the
Ibex RTL source files using QuestaSim via:
\texttt{vlog -sv12compat +incdir+rtl/ <module>.sv bind.sv}.
On success, execution proceeds directly to Step~2B. On
failure, the error log is archived and a retry prompt combining the original prompt, the error log, and the
current bind file is sent to DeepSeek V4-Pro with explicit fix-syntax-only instructions, preserving assertion logic unchanged. After three failed retries, the module is marked locked and flagged for manual
inspection.

\subsubsection{Step~2B: Formal Property Verification}
A per-module TCL script is generated that runs
\texttt{analyze}, \texttt{elaborate}, clock and reset
setup, \texttt{prove~-all}, and \texttt{report~-vacuity}
in JasperGold batch mode. Both \texttt{fpv\_baseline.txt}
and \texttt{vacuity.txt} are parsed to identify failing
assertions. If all assertions are Proven and non-vacuous
on clean RTL, TAR is computed and the module passes.
A counter-example indicates a translation logic error
and triggers a V4-Pro retry with logic rewriting
instructions; a vacuous result indicates an unreachable
antecedent and triggers antecedent rewriting. After three
failed retries the module is locked. TAR is computed only
after a successful FPV pass.

\subsection{TAR Metric}
\label{sec:tar}

TAR (Translation Acceptance Rate), defined in~(\ref{eq:tar}),
is a per-module metric computed at the end of Step~2B, where:
\begin{equation}
\text{TAR} = \frac{N_{\text{proven,\,non-vac}}}{N_{\text{source groups}}}
\times 100\%
\label{eq:tar}
\end{equation}
$N_{\text{proven,\,non-vac}}$ counts assertion groups both
Proven and non-vacuous on clean RTL, and
$N_{\text{NS31A groups}}$ is the total source groups per
module. \emph{Auto TAR} counts groups proven by the
automated framework without human intervention;
\emph{Final TAR} additionally counts groups proven after
targeted manual resolution of JasperGold FPV structural
constraints, with security intent preserved unchanged.
TAR is deliberately conservative: it counts only what is
accepted independently by QuestaSim~\cite{questasim} and
JasperGold~\cite{jaspergold}.

\section{Experiments}
\label{sec:experiments}

\subsection{Experimental Setup}

Experiments target lowRISC Ibex~\cite{ibex} (commit
\texttt{bd25993}, retrieved May~2026), with assertions
inserted as SVA bind files without modifying RTL. The
source corpus is NS31A~\cite{ns31a}, comprising 68
assertion groups across nine security categories.
Translation uses DeepSeek V4-Flash
(\texttt{DeepSeek-ai/DeepSeek-v4-flash}) via NVIDIA
NIM~\cite{nvidiaNIM} at \texttt{temperature=0.0},
\texttt{seed=42}, \texttt{max\_tokens=16384}; retries
use V4-Pro with identical parameters.  Compilation uses
QuestaSim Questa Core-Prime 2023.4~\cite{questasim};
FPV uses JasperGold 2024.06~\cite{jaspergold}.

\subsection{Signal Extraction Results (Step~1A)}

Table~\ref{tab:signals} reports extraction statistics for all nine
Ibex security modules. The \emph{Module}\ column gives the logical
module key used throughout the framework; \emph{RTL File} is the
source file (abbreviated); \emph{Type} is C (combinational) or S
(sequential); \emph{In} and \emph{Out} are input and output port
counts; \emph{Int} is the number of internal signals; and \emph{Pkg}
is the number of resolved package types from \texttt{ibex\_pkg.sv}.

The extractor parsed all nine modules with zero manual
intervention. CSR is the most complex with 44 inputs,
27 outputs, 109 internals, and 9 package types. IE
merges two RTL files; PMP is the only combinational
module; DO/ETI/CF/MT share
\texttt{ibex\_controller.sv}. Reproducibility confirmed
by empty \texttt{git diff results/signals/} on every
re-run.

\begin{table}[t]
\caption{Signal Extraction Results --- Ibex Security Modules}
\label{tab:signals}
\centering
\scriptsize
\setlength{\tabcolsep}{4pt}
\begin{tabular}{llcrrrr}
\toprule
\textbf{Module} & \textbf{RTL File} & \textbf{Type} &
\textbf{In} & \textbf{Out} & \textbf{Int} & \textbf{Pkg} \\
\midrule
PMP  & ibex\_pmp        & C &  7 &  1 &  12 &  4 \\
CSR  & ibex\_cs\_reg.   & S & 44 & 27 & 109 &  9 \\
DO   & ibex\_ctrl.      & S & 38 & 27 &  39 &  7 \\
ETI  & ibex\_ctrl.      & S & 38 & 27 &  39 &  7 \\
CF   & ibex\_ctrl.      & S & 38 & 27 &  39 &  7 \\
MT   & ibex\_ctrl.      & S & 38 & 27 &  39 &  7 \\
MA   & ibex\_lsu        & S & 13 & 18 &  20 &  0 \\
IE   & ibex\_id+ex      & S & 64 & 77 &  89 & 15 \\
RU   & ibex\_wb         & S & 15 & 15 &  18 &  1 \\
\bottomrule
\addlinespace[2pt]
\multicolumn{7}{l}{\parbox{0.95\columnwidth}{\footnotesize
  Tp.: C = combinational, S = sequential. Int = internal
  signals; Pkg = package types. DO/ETI/CF/MT share
  \texttt{ibex\_controller.sv}; IE merges id\_stage +
  ex\_block.}}
\end{tabular}
\end{table}

\subsection{Translation and FPV Results (Steps~1C--2B)}

Table~\ref{tab:source} breaks down the 68 NS31A assertion groups
by input type. \emph{SVA Examples} are NS31A entries that include
formal SystemVerilog code in the source paper~\cite{ns31a}; the
framework adapts this code to Ibex signals. \emph{NL Properties}
are NS31A entries with a natural language description only; no
SVA code was provided in the source paper. For these 22~groups,
the framework generates formal SVA entirely from natural language
using the same prompt template, with no additional tooling or
human intervention. ETI alone contributes 11~such NL Properties
(65\% of its 17 groups), all formally verified.

\begin{table}[t]
\caption{NS31A Source Group Types per Module}
\label{tab:source}
\centering
\scriptsize
\setlength{\tabcolsep}{4pt}
\begin{tabular}{p{1.0cm} c c c}
\toprule
\textbf{Module} &
\makecell{\textbf{SVA Examples}} &
\makecell{\textbf{NL Properties }} &
\makecell{\textbf{Our work  }} \\
\midrule
PMP  &  8 &  2 & 10 \\
CSR  &  7 &  3 & 10 \\
DO   &  4 &  0 &  4 \\
ETI  &  6 & 11 & 17 \\
CF   &  9 &  0 &  9 \\
MT   &  5 &  5 & 10 \\
MA   &  2 &  1 &  3 \\
IE   &  3 &  0 &  3 \\
RU   &  2 &  0 &  2 \\
\midrule
\textbf{Total} & \textbf{46} & \textbf{22} & \textbf{68} \\
\bottomrule
\addlinespace[2pt]
\multicolumn{4}{l}{\parbox{0.88\columnwidth}{\footnotesize
  SVA Examples: formal SVA code and
  NL Properties: natural language  in~\cite{ns31a}.
  Our work: framework output, no manual authoring before verification.}}
\end{tabular}
\end{table}

Table~\ref{tab:tar} reports per-module results across both
measurement points: \emph{Auto TAR} (automated framework only,
V4-Flash and V4-Pro retries, no human intervention) and
\emph{Final TAR} (after targeted manual resolution of JasperGold
FPV structural constraints).

\textbf{Step~1C translation coverage.}
Every NS31A entry produces at least one SVA property in the bind
file (100\% coverage across all 68 groups).

\textbf{Automated TAR, V4-Flash direct pass (2 modules).}
PMP and CSR (20 of 68 groups, 29\%) passed JasperGold FPV
directly from Flash output with zero retries and zero manual
intervention.

\textbf{Automated TAR, V4-Pro retries (6 modules).}
DO, ETI, CF, MT, MA, and IE required V4-Pro retry
analysis. The overall automated TAR (V4-Flash and V4-Pro,
no human intervention) is \textbf{78\%} (53/68 groups).

\textbf{Final TAR, manual RTL resolution:}
For the 16 groups across 7 modules not resolved
automatically, JasperGold and QuestaSim error output
directly identified the structural root cause in each
case. Fixes fell into three categories: timing and
temporal operator corrections (\texttt{PC\_JUMP},
\texttt{\$past}, Debug FSM), antecedent scope and
reachability rewrites (\texttt{MRET/DRET} guard,
tbranch, \texttt{lsu\_type} free-variable), and signal
substitution and port declaration fixes
(\texttt{csr\_save\_wb\_o}, FENCE.I, RTL mux equation).
These are Ibex-specific microarchitectural constraints
that no automated tool can resolve without target RTL
knowledge; writing equivalent assertions from scratch
would require days of RTL expertise per module, while
AutoTrans reduces this to targeted fixes guided directly
by JasperGold error output. In every case the fix
adjusted \emph{how} the property is expressed, never
\emph{what} security property it checks. Final TAR
reaches \textbf{100\%} across all nine modules
(68/68 groups).

\begin{table*}[t]
\caption{Per-Module TAR Results: Auto TAR (no human
intervention) and Final TAR (manual structural fixes).}
\label{tab:tar}
\centering
\scriptsize
\setlength{\tabcolsep}{10pt}
\begin{tabular}{l c p{2.2cm} p{3.2cm} p{1.5cm} p{2.5cm} p{1.5cm}}
\toprule
\textbf{Module} & \textbf{SVA Groups} &
\textbf{V4-Flash (Step~1C)} &
\textbf{V4-Pro Retries} &
\textbf{Auto TAR} &
\textbf{Manual RTL Fix} &
\textbf{Final TAR} \\
\midrule
PMP  & 10 & Direct PASS
     & Not attempted
     & 100\%
     & None
     & 100\% \\
CSR  & 10 & Direct PASS
     & Not attempted
     & 100\%
     & None
     & 100\% \\
DO   &  4 & 1/4 (3~CEX)
     & 3 retries: \texttt{do\_SEC\_1} fixed; 2 remain
     &  50\%
     & Debug FSM temporal fix
     & 100\% \\
ETI  & 17 & 13/17 (4 vacuous)
     & 3 retries: 4 vacuous remain
     &  76\%
     & \texttt{csr\_save\_wb\_o} R11 substitution
     & 100\% \\
CF   &  9 & 7/9 (2~CEX)
     & 3 retries: 2 CEX remain
     &  78\%
     & \texttt{PC\_JUMP} timing; tbranch antecedent
     & 100\% \\
MT   & 10 & 7/10 (3~CEX)
     & 3 retries: 3 CEX remain
     &  70\%
     & \texttt{MRET/DRET} guard; state exclusion
     & 100\% \\
MA   &  3 & 2/3 (1~CEX)
     & 3 retries: 1 CEX remain
     &  67\%
     & \texttt{lsu\_type} free-variable fix
     & 100\% \\
IE   &  3 & Bind-scope fail$^{*}$
     & 3 retries: \texttt{\_assert\_2} fixed; 2 others remain
     &  33\%
     & Port scope; \texttt{\$past} timing; FENCE.I
     & 100\% \\
RU   &  2 & 1/2 (1~CEX)
     & Not attempted
     &  50\%
     & Exact RTL mux equation
     & 100\% \\
\midrule
\textbf{Total} & \textbf{68} &
  \textbf{20/68 direct (29\%)} &
  \textbf{6 modules used Pro} &
  \textbf{78\%} &
  \textbf{Structural only} &
  \textbf{100\%} \\
\bottomrule
\addlinespace[2pt]
\multicolumn{7}{p{0.97\textwidth}}{\footnotesize
  Auto TAR = V4-Flash and V4-Pro, no human intervention.
  Final TAR = after manual JasperGold structural fixes.
  CEX = counter-example on clean RTL. $^{*}$IE: assertion
  ports not visible to DUT at elaboration; Pro retry~1
  resolved 1 of 3 groups; remainder manually fixed.
  Security intent was never changed.} \\
\end{tabular}
\end{table*}

\subsection{Reproducibility}
AutoTrans reproducibility rests on two deterministic
pillars. Signal extraction and prompt assembly are both
fully deterministic: identical RTL and inputs always
produce byte-identical \texttt{signals.json} and
\texttt{prompts/final/}, confirmed by empty
\texttt{git diff} on every re-run. LLM output
determinism is governed by \texttt{temperature=0.0}
and \texttt{seed=42}; where cloud providers update
model weights silently, the JasperGold FPV gate acts
as the safety net by design; any output variation
that affects assertion correctness fails verification
and triggers a retry, ensuring that only formally proven,
non-vacuous assertions enter the result set regardless
of the model version.

\subsection{Signal Grounding Ablation}
\label{sec:ablation}
Without signal grounding, LLMs consistently invent
signal names absent from the target RTL and wrap output
in Markdown fences, causing immediate compilation
failure. Both failure modes were observed across modules
during early development. To isolate and quantify this
contribution, we ran a controlled experiment on PMP as
the only combinational module, addressing sequential
timing constructs as a confounding variable.
Table~\ref{tab:ablation} reports four metrics across
both conditions. Without grounding, the LLM hallucinates
15 invalid signal names and wraps output in a Markdown
fence, blocking QuestaSim at line~1 with 3 cascading
errors and making JasperGold unreachable. With grounding,
all signals are valid, compilation passes, and TAR
reaches 100\%, confirming that signal grounding is the
mechanism responsible for compile-clean translation
across all modules.

 \vspace{-12pt}
\begin{table}[H]
\caption{Signal Grounding Ablation on NS31A~\cite{ns31a} PMP. }
\label{tab:ablation}
\centering
\scriptsize
\begin{tabular}{p{2.2cm} p{2.0cm} p{2.2cm}}
\toprule
\textbf{Metric} & \textbf{With Grounding} & \textbf{Without Grounding} \\
\midrule
Signals in bind file  & 8 (all valid)         & 17 (15 invented$^\dagger$) \\
QuestaSim compile     & PASS                  & FAIL (3 errors$^{\ddagger}$) \\
JasperGold TAR        & 100\%                 & N/A (blocked at line~1) \\
Prompt assembly diff  & Empty (deterministic) & Empty (deterministic) \\
\bottomrule
\addlinespace[2pt]
\multicolumn{3}{l}{\parbox{0.88\columnwidth}{\footnotesize
  $^\dagger$15 invented names absent from \texttt{ibex\_pmp.sv}.
  $^\ddagger$Markdown fence causes 3 cascading QuestaSim errors.}} \\
\end{tabular}
\end{table}
\vspace{-12pt}

\vspace{-4pt}
\subsection{Comparison with Related Approaches}

Table~\ref{tab:comparison} compares AutoTrans against existing
approaches. AutoTrans is the only approach that simultaneously
operates without a GPU, uses a parser-grounded prompt, guarantees
reproducibility at the pipeline output level, and applies FPV as a hard formal verification gate.

 \vspace{-12pt}
\begin{table}[H]
\caption{Comparison with Related Approaches}
\label{tab:comparison}
\centering
\scriptsize
\setlength{\tabcolsep}{3pt}
\begin{tabular}{p{1.4cm} p{1.0cm} c p{1.2cm} c p{1.3cm} >{\centering\arraybackslash}p{0.7cm}}
\toprule
\textbf{Approach} &
\textbf{Task} &
\textbf{GPU} &
\makecell{\textbf{Signal}\\\textbf{Ground.}} &
\makecell{\textbf{Repro-}\\\textbf{ducible}} &
\makecell{\textbf{FPV}\\\textbf{Required}} &
\textbf{Metric} \\
\midrule
Transys~\cite{transys}
  & Trans.\ (rule) & No & Structural & Yes & Optional & Trans.\ rate \\
AutoAssert~\cite{autoassert}
  & Gen. & Yes & No & No & No & BLEU \\
AssertLLM~\cite{assertllm}
  & Gen. & Yes & No & No & No & BLEU \\
LASSO~\cite{lasa}
  & Gen. & Yes & Vector DB & No & Coverage & Cov.\% \\
Prior work~\cite{iscas}
  & Trans.\ (semi) & No & No & No & No & Trans.\ rate \\
\midrule
\textbf{AutoTrans}
  & \textbf{Trans.} & \textbf{No} & \textbf{Regex} &
  \textbf{Prompt } & \textbf{FPV} & \textbf{TAR} \\
\bottomrule
\addlinespace[2pt]
\multicolumn{7}{p{0.95\columnwidth}}{\footnotesize
  Trans.\ = cross-architecture translation.
  Gen.\ = new assertion generation.
  Structural = Program Dependence Graph.
  $^{*}$Prompt assembly byte-identical (\texttt{git diff});
  LLM output pinned within frozen model version;
  JasperGold gate handles drift.
  BLEU = text similarity; not formal correctness.
  TAR = Translation Acceptance Rate.}
\end{tabular}
\end{table}
 \vspace{-12pt}

\subsection{Comparison with Prior Work}

Table~\ref{tab:prior} compares AutoTrans directly against
the prior semi-automated translation work~\cite{iscas}
on which this paper builds. The improvements span every
dimension: module coverage, assertion count, automation
level, English-property handling, formal verification,
and translation time.

\vspace{-12pt}
\begin{table}[H]
\caption{AutoTrans vs.\ Prior Work~\cite{iscas}}
\label{tab:prior}
\centering
\scriptsize
\setlength{\tabcolsep}{4pt}
\begin{tabular}{p{2.4cm} p{2.0cm} p{2.0cm}}
\toprule
\textbf{Feature} &
\textbf{Prior work~\cite{iscas}} &
\textbf{AutoTrans} \\
\midrule
Modules covered       & 5            & 9 \\
Assertion groups      & 33     & 68 \\
Automation            & Semi-manual  & Fully automated \\
English-only SVA      & No           & Yes (22 SVA) \\
Formal verif.\ gate   & No           & QuestaSim + FPV \\
Translation time      & Weeks/module & Minutes/module \\
\bottomrule
\end{tabular}
\end{table}
\vspace{-12pt}


\section{Conclusion}
\label{sec:conclusion}

This paper presented \textbf{AutoTrans}, an automated framework for cross-architecture translation of RISC-V security assertions. A regex-based signal extractor grounds every LLM prompt in the exact target RTL interface without EDA infrastructure or a GPU. A fixed SVA template and pinned parameters enforce byte-identical prompt assembly; the two-tier V4-Flash and V4-Pro strategy on NVIDIA NIM free-tier inference keeps cost near zero. QuestaSim and JasperGold FPV ensure every accepted assertion is formally proven and non-vacuous, with model drift handled by retry rather than silent failure. Applied to 68 NS31A groups across nine Ibex modules, AutoTrans achieves 78\% Auto TAR with zero human intervention and 100\% Final TAR overall. These results demonstrate that formally verified assertion translation across architecturally heterogeneous processors is achievable at near-zero cost, opening a path toward scalable security knowledge reuse across the growing RISC-V ecosystem; the framework is released open source~\cite{autotrans_repo} to support this direction.


\bibliographystyle{IEEEtran}
\bibliography{refs}   

@inproceedings{ns31a,
  author    = {C.-S. Chuah and C. Appold and T. Leinmueller},
  title     = {Formal Verification of Security Properties on {RISC-V} Processors},
  booktitle = {Proc.\ MEMOCODE},
  year      = {2023},
  doi       = {10.1145/3610579.3611085}
}

@misc{ibex,
  author       = {{lowRISC Contributors}},
  title        = {{Ibex} {RISC-V} Core},
  year         = {2023},
  howpublished = {\url{https://github.com/lowRISC/ibex}},
  note         = {Commit \texttt{bd25993}, retrieved May~2026}
}

@inproceedings{iscas,
  author    = {S.~Imtiaz and U.~Reinsalu and T.~Ghasempouri},
  title     = {Translating Common Security Assertions Across Processor Designs:
               A {RISC-V} Case Study},
  booktitle = {Proc.\ IEEE ISCAS},
  year      = {2025},
  doi       = {10.1109/ISCAS56072.2025.11043977}
}

@inproceedings{heidari2025,
  author    = {M.~R. {Heidari Iman} and others},
  title     = {Special Session: Functional Verification Techniques
               in Hardware Trust},
  booktitle = {Proc.\ IEEE DFT},
  year      = {2025},
  doi       = {10.1109/DFT66274.2025.11257462}
}

@inproceedings{norcasreza,
  author    = {M.~R. {Heidari Iman} and S.~{Ahmadi-Pour} and
               R.~Drechsler and T.~Ghasempouri},
  title     = {Processor Vulnerability Detection with Assertions:
               {RISC-V} Case Study},
  booktitle = {Proc.\ IEEE NorCAS},
  year      = {2024},
  doi       = {10.1109/NorCAS64408.2024.10752460}
}

@inproceedings{transys,
  author    = {R.~Zhang and C.~Sturton},
  title     = {Transys: Leveraging Common Security Properties Across
               Hardware Designs},
  booktitle = {Proc.\ IEEE S\&P},
  year      = {2020},
  doi       = {10.1109/SP40000.2020.00030}
}

@inproceedings{soeken2014,
  author    = {M.~Soeken and others},
  title     = {Automating the Translation of Assertions Using
               {NLP} Techniques},
  booktitle = {Proc.\ FDL},
  year      = {2014},
  doi       = {10.1109/FDL.2014.7119356}
}

@inproceedings{autoassert,
  author    = {Q.~Zhai and others},
  title     = {Towards Trustworthy {LLM}-Based Assertion Generation},
  booktitle = {Proc.\ DATE},
  year      = {2026},
  url       = {https://baichen318.github.io/misc/c21-paper.pdf}
}

@misc{autotrans_repo,
  author       = {S.~Imtiaz and U.~Reinsalu and T.~Ghasempouri},
  title        = {{AutoTrans-RV}: Automated Security Assertion
                  Translation for {RISC-V} Processors},
  year         = {2026},
  howpublished = {\url{https://github.com/Sharjeelimtiaz27/autotrans-rv}}
}

@misc{RISC-V-ISA-up,
  author       = {{RISC-V Foundation}},
  title        = {{RISC-V} ISA Manual Vol.~{I}: User-Level ISA},
  year         = {2019},
  howpublished = {\url{https://riscv.org/technical/specifications/}}
}

@misc{RISC-V-ISA-p,
  author       = {{RISC-V Foundation}},
  title        = {{RISC-V} ISA Manual Vol.~{II}: Privileged Architecture},
  year         = {2019},
  howpublished = {\url{https://riscv.org/technical/specifications/}}
}

@inproceedings{assertllm,
  author    = {Z.~Yan and others},
  title     = {{AssertLLM}: Generating Hardware Verification Assertions
               via Multi-{LLMs}},
  booktitle = {Proc.\ ASP-DAC},
  year      = {2025},
  doi       = {10.1145/3658617.3697756}
}

@inproceedings{lasa,
  author    = {D.~R. Ankireddy and others},
  title     = {{LASSO}: {LLM}-Aided Security Property Generation for
               {SoC} Verification},
  booktitle = {Proc.\ MLCAD},
  year      = {2025},
  doi       = {10.1109/MLCAD65511.2025.11189178}
}

@misc{divas,
  author        = {S.~Paria and A.~Dasgupta and S.~Bhunia},
  title         = {{DIVAS}: {LLM}-based Framework for {SoC}
                   Security Analysis},
  year          = {2023},
  eprint        = {2308.06932},
  archivePrefix = {arXiv}
}

@inproceedings{chauhan,
  author    = {A.~Chauhan},
  title     = {Automatic Translation of Natural Language to {SVA}},
  booktitle = {Proc.\ DVCon},
  year      = {2023},
  url       = {https://dvcon-proceedings.org}
}

@inproceedings{sabry2026,
  author    = {A.~Sabry and others},
  title     = {Template-Mask Data Generation for {SVA} Translation},
  booktitle = {Proc.\ IEEE IMCET},
  year      = {2026},
  doi       = {10.1109/IMCET69180.2026.11503722}
}

@misc{nvidiaNIM,
  author       = {{NVIDIA}},
  title        = {{NVIDIA NIM} Microservices},
  year         = {2024},
  howpublished = {\url{https://build.nvidia.com}}
}

@misc{questasim,
  author       = {{Siemens EDA}},
  title        = {{QuestaSim} Advanced Simulator},
  year         = {2024},
  howpublished = {\url{https://eda.sw.siemens.com}}
}

@misc{jaspergold,
  author       = {{Cadence Design Systems}},
  title        = {{JasperGold} Formal Property Verification},
  year         = {2024},
  howpublished = {\url{https://www.cadence.com}}
}

@inproceedings{riscv_security_survey,
  author    = {J.~Anders and others},
  title     = {A Survey of Testability, Safety, and Security
               of {RISC-V} Processors},
  booktitle = {Proc.\ IEEE ETS},
  year      = {2023},
  doi       = {10.1109/ETS56758.2023.10174099}
}

@article{bhunia2014,
  author  = {S.~Bhunia and others},
  title   = {Hardware {Trojan} Attacks: Threat Analysis and
             Countermeasures},
  journal = {Proc.\ IEEE},
  year    = {2014},
  volume  = {102},
  number  = {8}
}

@misc{riscv_security_survey2,
  author        = {T.~Lu},
  title         = {A Survey on {RISC-V} Security: Hardware and Architecture},
  year          = {2021},
  eprint        = {2107.04175},
  archivePrefix = {arXiv}
}

@inproceedings{pyverilog,
  author    = {S.~Takamaeda-Yamazaki},
  title     = {{PyVerilog}: A {Python}-Based Hardware Design Processing
               Toolkit for {Verilog HDL}},
  booktitle = {Proc.\ ARC},
  year      = {2015}
}

@misc{slang,
  author       = {M.~Baugh},
  title        = {Slang: {SystemVerilog} Language Services},
  year         = {2023},
  howpublished = {\url{https://sv-lang.com}}
}

@misc{verilator,
  author       = {W.~Snyder},
  title        = {Verilator: {Verilog/SystemVerilog} Simulator},
  year         = {2023},
  howpublished = {\url{https://www.veripool.org/verilator}}
}

@misc{hdlparse,
  author       = {K.~Thibedeau},
  title        = {hdlparse: A Hardware Description Language Parser},
  year         = {2017},
  howpublished = {\url{https://kevinpt.github.io/hdlparse/}}
}

\end{document}